\documentclass[traditabstract]{aa}  
\usepackage{graphicx, xcolor}
\usepackage{longtable,lscape}
\usepackage{dcolumn}
\usepackage{txfonts}
\usepackage[authoryear]{natbib} 
\bibpunct{(}{)}{,}{a}{}{,} % citation like A&A style
\begin{document}
   \title{The Quintuplet Cluster}

   \subtitle{III. Hertzsprung-Russell diagram and cluster age}

   \author{A. Liermann \inst{1} \and
           W.-R. Hamann \inst{2} \and
           L. M. Oskinova \inst{2} 
          }

   \institute{Max-Planck Institut f\"ur Radioastronomie, 53121 Bonn,
              Germany \and 
              Universit\"at Potsdam, Institut f\"ur Physik und Astronomie,
              14476 Potsdam, Germany 
             }

   \date{Received September 15, 1996; accepted March 16, 1997}

   \titlerunning{The Quintuplet Cluster. III.}

\abstract{The Quintuplet, one of three massive stellar clusters in the
Galactic center (GC), is located about 30\,pc in projection from
Sagittarius A$^*$. We aim at the construction of the Hertzsprung-Russell
diagram (HRD) of the cluster to study its evolution and to constrain its
star-formation history. For this purpose we use the most complete
spectral catalog of the Quintuplet stars.
Based on the $K$-band spectra we determine stellar temperatures
and luminosities for all stars in the catalog under the assumption of a
uniform reddening towards the cluster.  We find two groups in the
resulting HRD: early-type OB stars and late-type KM stars, well
separated from each other. By comparison with Geneva
stellar evolution models we derive initial masses exceeding
8\,$M_\odot$ for the OB stars. In the HRD these stars 
are located along
an isochrone corresponding to an age of about 4\,Myr. This confirms
previous considerations, where a similar age estimate was based on the
presence of evolved Wolf-Rayet stars in the cluster.
We derive number ratios for the various spectral subtype groups
(e.\,g. $N_\mathrm{WR}/N_\mathrm{O}$, $N_\mathrm{WC}/N_\mathrm{WN}$) and
compare them with predictions of population synthesis models. We find
that an instantaneous burst of star formation at about 3.3 to 3.6\,Myr
ago is the most likely scenario to form the Quintuplet cluster.
Furthermore, we apply a mass-luminosity relation to construct the
initial mass function (IMF) of the cluster. 
We find indications for a slightly top-heavy IMF.
The late-type stars in the LHO
catalog are red giant branch (RGB) stars or red supergiants (RSGs)
according to their spectral signatures. Under the assumption that they
are located at about the distance of the Galactic center we can derive
their luminosities. The comparison with stellar evolution models reveals
that the initial masses of these stars are lower than 15\,$M_\odot$
implying that they needed about 15\,Myr (RSG) or even more than 30\,Myr (RGB) to
evolve into their present stage. It might be suspected that these
late-type stars do not physically belong to the Quintuplet cluster.
Indeed, most of them disqualify as cluster members because their radial
velocities differ too much from the cluster average. Nevertheless, five
of the brightest RGB/RSG stars from the LHO catalog share the
mean radial velocity of the Quintuplet, and thus remain highly suspect
for being gravitationally bound members. If so, this would challenge the
cluster formation and evolution scenario.
}

   \maketitle
%
%________________________________________________________________

\section{Introduction}
Stellar clusters are unique objects to study stellar populations and their 
evolution, as they are supposed to represent a single coeval
  stellar population. 
The Quintuplet cluster is one of the massive young stellar clusters in the
Galactic center region. Located at about 30\,pc projected distance
from the Galactic Center, its age has been estimated to be about
4\,Myr\,\citep{FMM99} 
from the evolved massive stars in their Wolf-Rayet (WR) phase, 
i.e. stars displaying CNO-processed (WN stars) or helium
burning products (WC stars) in their spectra. The
WN stars were analyzed in detail  
by \citet{Liermann+2010} who found from comparison with stellar
evolution models ages of about 3\,Myr that agree quite well with the
previously assumed cluster age. 

Despite the young cluster age, \citet{GMM90} reported the presence of
one red supergiant (RSG) in the cluster. A few other Galactic
clusters are known to show the coexistence of evolved stars like WR
stars and RSGs, e.\,g. Westerlund\,1 \citep{Crowther+2006}, but it seems quite puzzling for a
young cluster like the Quintuplet.
\citet{Liermann+2009} presented the so far most  complete spectral
catalog of the Quintuplet stars. From this catalog the 
Hertzsprung-Russell diagram (HRD) can be constructed, and the results can
be compared to stellar evolution models and isochrones. This will allow to
put more stringent constrains on the age of the Quintuplet
cluster. The comparison
of number ratios with population synthesis models will help to
confine the cluster age further and 
allows a statement about the formation history to be made. Especially,
the apparent existence of RSGs in the cluster needs to be addressed. 

The paper is organized as follows. In Sect.\,\ref{sec:obs} we introduce
the spectra and the sample of stars, Sect.\,\ref{sec:analysis} explains the
methods to derive a cluster HRD. Finally, the results
are discussed in the context of stellar and cluster evolution in
Sect.\,\ref{sec:results}. 

\section{Observations}
\label{sec:obs}
The analysis in this paper is based on observations of the Quintuplet
cluster with the ESO Very Large Telescope's integral-field
spectrograph SINFONI, that cover $K$-band spectra with a resolution of
about $R \approx 4000$, complete to a photometric magnitude of about
$K_\mathrm{s} = 13$\,mag. The data were published in a catalog by
\citet[hereafter LHO catalog]{Liermann+2009}, listing 85 early-type
and 62 late-type stars with their spectral classification and
individual radial velocity ($RV$). A field of view of about
40\,\arcsec $\times$ 40\,\arcsec is covered in total.  Please refer to
the LHO catalog for more detailed information about observational
strategy and data reduction.

\section{Analysis}
\label{sec:analysis}
\subsection{Sample stars}
The LHO catalog gives a mean cluster radial velocity of
$RV = 113 \pm 37$\,km\,s$^{-1}$ and a first approximation for
  the velocity dispersion $\sigma = 17\,$km\,s$^{-1}$ within the  
Quintuplet cluster, under the assumption of a virialized cluster.
The cluster membership was rejected for stars with a 
radial velocity that
differs from the cluster mean by more than 3\,$\sigma$
($\pm$ 51\,km\,s$^{-1}$). However, field stars in the vicinity of the
Quintuplet may participate in a similar Galactocentric rotation and thus
show a similar $RV$, albeit not being bound to the cluster. Moreover, $RV$s
can be affected by the orbital motion for stars in close binary systems.
In these cases the periodic
variation of the $RV$s can only be detected from time series of
spectra, which are not (yet) available for the Quintuplet stars.

%________________________________________________________________
\begin{figure}
         \center{
         \includegraphics[width=\columnwidth]{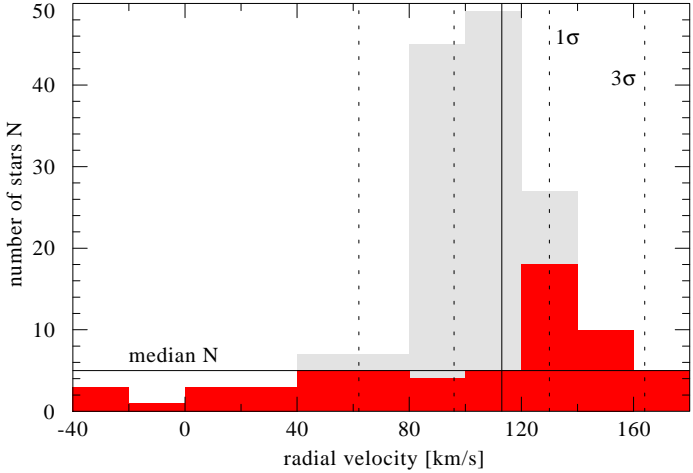}}
      \caption{Histogram of the distribution of $RV$s of all LHO
        stars (light gray). Overplotted is the distribution for the
        late-type KM
        stars (dark/red). Solid and dashed lines indicate the cluster $RV$ of
        113\,km s$^{-1}$ and the 1\,$\sigma$ and 3\,$\sigma$ intervals, respectively.}
         \label{fig:RV-cluster-KM}
\end{figure}
\begin{figure}
         \center{
      \includegraphics[width=\columnwidth]{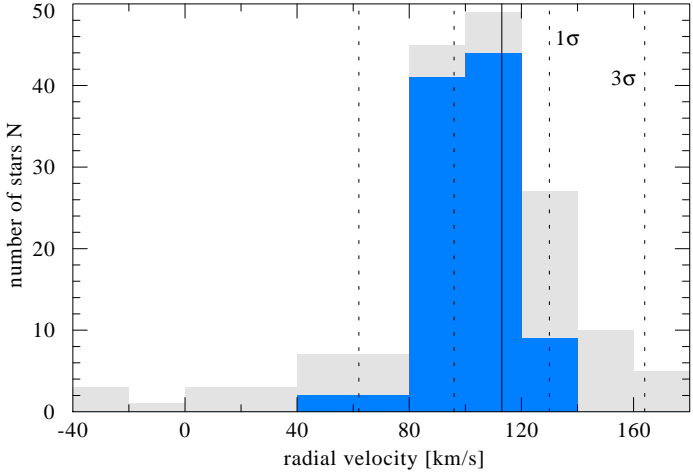}}
      \caption{Same as Fig.\,\ref{fig:RV-cluster-KM}, but overplotted with the
      $RV$ distribution of early-type OB stars (dark/blue). Both distributions peak
      within the intervals assumed for cluster membership, albeit more
      pronounced for the early-type stars.} 
      \label{fig:RV-cluster-OB}
\end{figure}
%________________________________________________________________
%
%
%________________________________________________________________
   \begin{figure}
   \centering
   \includegraphics[width=\columnwidth]{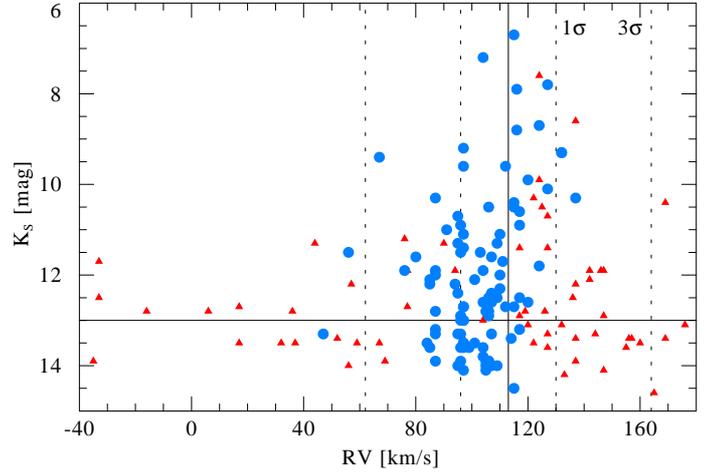}
   \caption{Distribution of $K_\textrm{s}$ magnitudes over radial 
   velocity for all stars in the LHO catalog (circles for
   early-type stars, triangles for late types). The
   completeness limit at $K_\textrm{s}=13\,$mag is indicated by
   the horizontal line, while the vertical lines refer to the
   different $RV$ intervals as in Fig.\,\ref{fig:RV-cluster-KM}.}
               \label{fig:RV-Kmag}
   \end{figure}
%________________________________________________________________
%

The number distributions of the radial velocities of the late-type
and early-type stars are shown in Figs.\,\ref{fig:RV-cluster-KM} and
\,\ref{fig:RV-cluster-OB}.
While we find a very pronounced peak in the distribution for the
early-type stars in the $RV$ range assumed to indicate cluster
membership, the late-type KM stars seem to be distributed more 
homogeneously with a somewhat lower peak within the 3$\sigma$
interval around the mean cluster $RV$. 
To estimate the number of stars, late-type stars in particular,
along a line-of-sight towards the Galactic center, we would
need to observe control fields around the Quintuplet. The lack of those requires
a more careful interpretation of Fig.\,\ref{fig:RV-cluster-KM}: from
the rather flat distribution, we could assume a median number of five
stars per velocity bin to possibly be field stars.

Additionally, the question of foreground stars has to be
regarded under the aspect that the high visual extinction towards the
Galactic center is overcome in the near-IR range. This in return makes
it more difficult to distinguish between foreground and 
background objects relative to the cluster. As can be seen from
Fig.\,\ref{fig:RV-Kmag} there is no clear distinction in magnitude between
bright foreground or faint background objects,  
but a wide distribution of the $K$-band magnitudes
over the $RV$ range. The photometric completeness of the LHO catalog
at $K_\textrm{s}=13$\,mag hardly alters the distribution. 

We set the following criteria to assess the possible cluster
membership:
stars within the 1\,$\sigma$ interval of the cluster $RV$ and brighter
than the photometric completeness limit, i.\,e. $K_\textrm{s}< 13$\,mag,
will be considered as very likely cluster members. We will refer to this
sample of stars as the  {\it corrected sample} in the following in
comparison to the complete LHO sample. \footnote{The criteria are
slightly stricter than the previous 3\,$\sigma$ cluster $RV$, also
applied in the LHO catalog, but are meant to compensate to some degree
the lack of time series of spectra to exclude binary effects and the
lack of control fields to estimate the number of field stars
properly.}

\subsection{Stellar luminosities and the HRD}
The HRD is a useful tool to characterize a
stellar cluster; to establish it we need to know the effective temperatures and
stellar luminosities for the Quintuplet stars.

The effective temperature is defined via the Stefan-Boltzmann law, and
thus follow from the stellar luminosity and radius. This reference
radius is a matter of definition. Usually one refers the stellar
radius to a Rosseland optical depth of $2/3$. In stars with a well
defined photosphere the precise definition does not matter. In WR
atmospheres, however, $\tau = 2/3$ is 
often reached already in moving layers, i.\,e. in the wind. Therefore it
became standard to define the ``stellar radius'' $R_\ast$ of
WR stars at a Rosseland optical depth of 20. In most models, this point
lies at subsonic velocities, i.e.\ in nearly hydrostatic layers. The
effective temperature related to $R_\ast$ is termed the ``stellar
temperature'' $T_\ast$, and this value is quoted for the WN stars as
the result from the spectral analyses from \citet{Liermann+2010}.

The motivation for using the radius of the hydrostatic part is the hope
that $R_\ast$ can be identified with the stellar radius from stellar
evolution models. However, our empirical $R_\ast$ is in fact based on an
inward extrapolation of the stellar wind's velocity law into optically
thick, i.e. un-observable layers. In the case of our Quintuplet WN
stars, this is not critical because $R_\ast$ is only slightly smaller than
the radius where $\tau = 2/3$ -- only in very thick WR winds, the
difference becomes significant. Nevertheless, there is evidence that most
WR stars have very extended subphotospheric layers which make $R_\ast$
much bigger (and thus $T_\ast$ much lower) than current models of the
stellar structure predict.

For the WC stars in the Quintuplet the analysis is in
progress. All of them are classified as WC8-9 and their majority
produces dust \citep{vdH2006, Liermann+2009}. This dust production is 
considered to arise from colliding winds in high-mass binaries, as
has been confirmed already for some of the Quintuplet WC stars by
\citet[``pinwheel stars'']{Tuthill+2006}.
Therefore, the WC stars will be excluded from the present
considerations of the cluster HRD, as they need to be analyzed and
discussed with respect to the special scenarios of binary evolution.

For the late-type KM stars, we determine the effective temperatures
($T_\mathrm{eff}$) according to their spectral classification in the LHO
catalog. Those were based on the equivalent widths measured from the
first overtone CO absorption band at 2.3\,$\mu$m
\citep[see][]{Gonzalez-Fernandez+2008}. The $T_\mathrm{eff}$ are then
applied to obtain the bolometric correction for the $K$-band magnitude,
BC$_K$, adopting the relation 
\begin{equation} 
\textrm{BC}_K = 5.574 - 0.7589\,(T_\mathrm{eff}/1000\textrm{K})\,,
\end{equation}
derived by \citet{Levesque+2005} for red supergiants. We apply the same 
relation for our red giant branch stars (RGBs) as well, thus
neglecting effects of the luminosity class on the spectral energy
distribution. 

The effective temperatures for the early-type OB type stars were
determined from their spectral (sub-)types according to the
spectral-type--temperature calibration of \citet[Tables\,2,3,5, and
  6]{Martins+2005} and applied for their bolometric corrections
\begin{equation}
\textrm{BC}_V = 27.58 - 6.80\,\log{(T_\mathrm{eff}/\textrm{K})}\,.
\end{equation}
These BCs are valid only for the visual spectral
range as indicated by the index and a correcting term for the $K$-band
BCs is necessary.
From a previous model analysis of O stars \citep{Oskinova+2006} we
can estimate an average $M_V-M_K = (-0.98 \pm 0.04)$\,mag for their
sample stars. That sample contained spectral types O3 to O7 with
luminosity classes I to V, therefore the term is considered to be robust and
applicable for the correction $\textrm{BC}_K = \textrm{BC}_V - (M_V-M_K)$.

Absolute stellar magnitudes $M_K$ were derived from the
$K_\textrm{s}$ given in the LHO catalog with the mean cluster
extinction $A_K= 3.1 \pm 0.5$\,mag derived from the analyses of the WN
stars \citep{Liermann+2010} together with the corresponding BCs. For the
cluster distance we adopt the distance to the GC of 8\,kpc
\citep{Reid1993}. 

\begin{figure}
         \center{
\includegraphics[width=0.9\columnwidth]{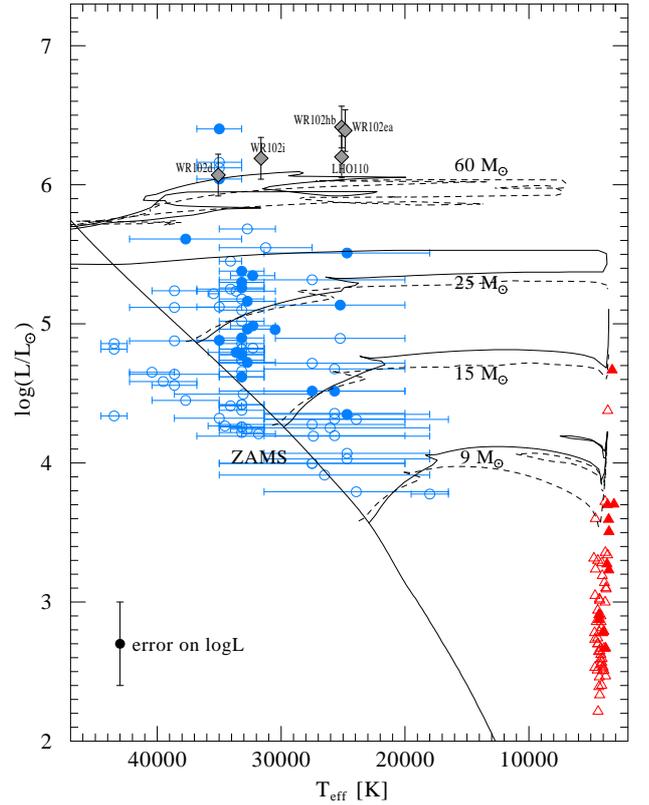}}
         \caption{Hertzsprung-Russell diagram of the Quintuplet. Circles
         (blue) represent the early-type OB stars, triangles (red) the  
       late-type KM stars, filled symbols refer to stars of the corrected
       sample (see text). The ZAMS and stellar evolution tracks with
       rotation (solid lines) and without rotation (dashed lines) for
       different initial masses are from \citet{Meynet+Maeder2003}.}  
     \label{fig:cluster-hrd+tracks}
\end{figure}

The absolute magnitudes are transformed to stellar luminosities 
\begin{equation}
\log{\left( \frac{L}{L_\odot} \right)} = - 0.4\,(M_K - \textrm{BC}_K -
M_{\mathrm{bol,}\odot}) \,, 
\end{equation}
with $M_{\mathrm{bol,}\odot} = 4.74$\,mag for the bolometric luminosity of
the Sun.
The resulting HRD of the Quintuplet cluster is shown in
Fig.\,\ref{fig:cluster-hrd+tracks}. 

%________________________________________________________________
%________________________________________________________________

\vspace{.2cm}
\noindent
The errors on the temperature and luminosity of the WN stars are taken
from \citet{Liermann+2010}. 

For the late-type KM stars, the error on $T_\mathrm{eff}$ is of the order of 
200\,K \citep{Gonzalez-Fernandez+2008}. For those stars the error bars
vanish within the stars' symbol sizes. 

For the early-type OB stars the error on $T_\mathrm{eff}$ reflects the
range in the spectral classification in the LHO catalog as an upper
and lower temperature according to the spectral-type--temperature
calibration. This error is usually quite large and could be 
minimized by tailored modeling of the individual stars, which is out of scope for the work presented here.

For both late- and early-type stars we estimate the error on the
luminosity from the combined errors of the $K$-band magnitude,
extinction, and bolometric corrections. 
The error is dominated by the error of the mean cluster
extinction determined from the WN star analysis (0.5\,mag).
The error on the $K$-band magnitudes from the LHO catalog is given
with 0.2\,mag.
For the BCs \citet{Levesque+2005} list an error of 0.01\,mag for
late-type stars and \citet{Martins+2005} give 0.05\,mag for early-type
stars. We had to correct the BCs for the early-type stars from the
visual to the $K$-band with a correcting term for with we estimate an
error of 0.04\,mag (see above). In total, the combined error on the luminosity 
amounts to about 0.3\,dex for both the late- and early-type stars.  

%________________________________________________________________
%
\begin{figure}
         \center{
\includegraphics[width=0.9\columnwidth]{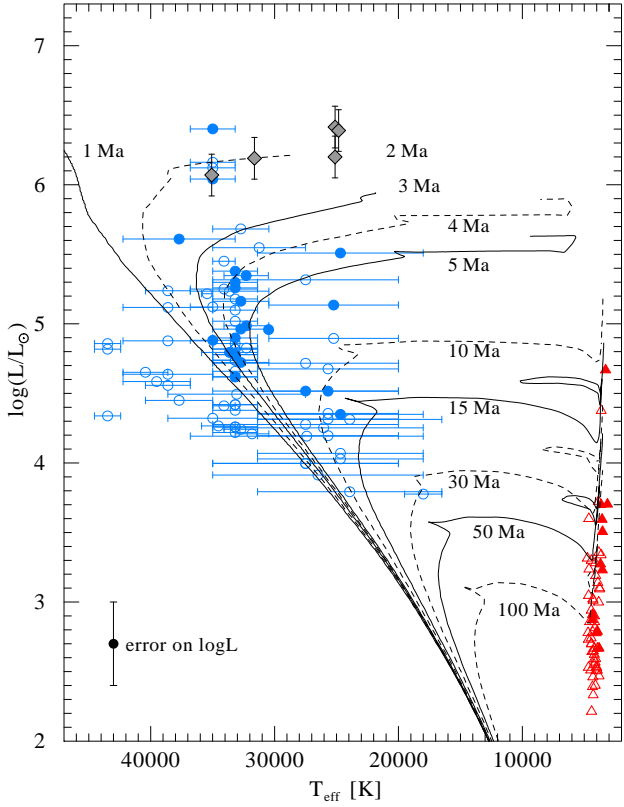}}
     \caption{Same as Fig.\,\ref{fig:cluster-hrd+tracks} but with 
     theoretical isochrones for different cluster ages.
     The isochrones, shown with alternating line styles, were constructed
     by \citet{Lejeune+2001}. 
     The majority of the population of OB stars in the Quintuplet follows the 4\,Myr
     isochrone, while most KM stars would need more than 30\,Myr to evolve}. 
      \label{fig:cluster-hrd+isochrones}
\end{figure}
\begin{figure}
         \center{
\includegraphics[width=0.9\columnwidth]{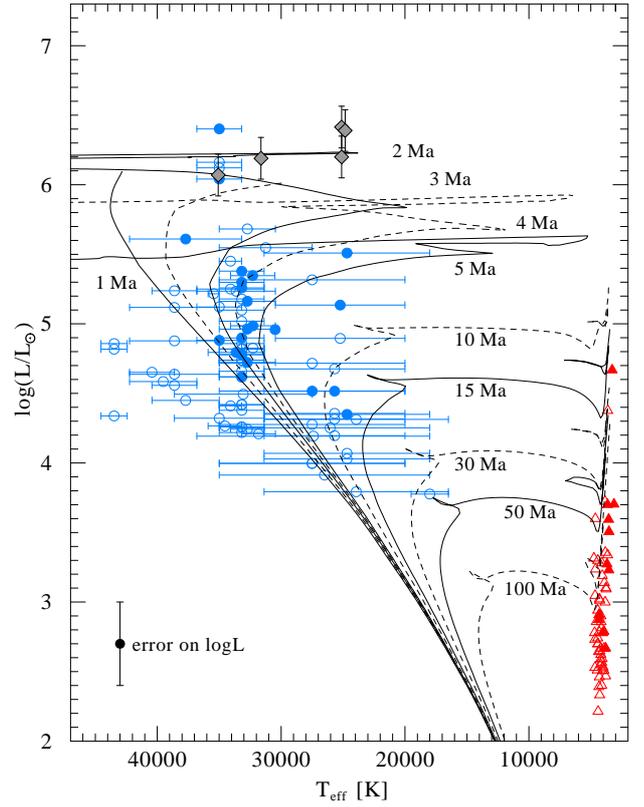}}
\caption{Same as Fig.\,\ref{fig:cluster-hrd+isochrones} but with
  theoretical isochrones based on \citet{Girardi+2002}. Again, the majority of
  OB stars falls on the 4\,Myr isochrone, while  most KM stars
    evolve on timescales with more than 30\,Myr}.
      \label{fig:cluster-hrd+isochrones2}
\end{figure}
%________________________________________________________________
%

\section{Discussion}
\label{sec:results}
\subsection{Stellar evolution and cluster age}
As can be clearly seen from Fig.\,\ref{fig:cluster-hrd+tracks}, two
distinct groups are found in the HRD. The OB stars are located
more or less along a main sequence, while the KM stars 
line up in the low-temperature regime with $T_\mathrm{eff} < 5000$\,K
over an extended luminosity range.
It seems that a number of OB stars are found to the left of 
the zero-age main sequence (ZAMS). On one hand, this can be attributed 
to the rather large error bars in temperature, i.e. uncertainty of the star's spectral type, and luminosity. However, 
it might also reflect the fact that the assumed homogeneous reddening 
towards the cluster is not correct. And with the presence of dusty WC stars an 
additional intrinsic reddening might have to be considered for the different 
regions of the cluster.

Stellar evolution tracks including the effects of rotation (solid lines)
and without rotation (dashed lines) for initial masses of 9, 15, 25 and
60\,$M_\odot$ \citep{Meynet+Maeder2003} show that the OB stars
are massive stars in the classical definition,
i.\,e. $M_\mathrm{init} > 8\,M_\odot$. We omit the tracks for 
higher masses as they overlap with the track for 60\,$M_\odot$ in case of 
the 85\,$M_\odot$ track or don't extend into the cooler temperature range 
in case of the 120\,$M_\odot$ track.  
The difference in stellar evolution between models with rotation and
without is stronger from 25\,$M_\odot$ upwards, i.\,e. the high-mass
stars potentially evolving to WR stars, and is less important for the
majority of stars in the current sample.

%________________________________________________________________
\begin{table*}[!ht]
  \caption{Number ratios for the Quintuplet stars in comparison to
    other Galactic center clusters.  
  } 
  \label{tab:number-ratios}
  \begin{center}
  \begin{tabular}{lcr@{:}lr@{:}lr@{:}lr@{:}lr@{:}lr@{:}l} \hline \hline
  \noalign{\smallskip}
  & $N_\mathrm{WR}$ & $N_\mathrm{WR}$ & $N_\mathrm{O}$ &
  $N_\mathrm{WC}$ & $N_\mathrm{WN}$ & 
  $N_\mathrm{RSG}$  & $N_\mathrm{WR}$ & $N_\mathrm{WNE}$&
  $N_\mathrm{WNL}$  & $N_\mathrm{WC, binary}$ & $N_\mathrm{WC}$ & 
  $N_\mathrm{WN, binary}$ & $N_\mathrm{WN}$\\    
  \noalign{\smallskip} \hline
  \noalign{\smallskip}
 Quintuplet$^*$    &21& -&- &13&~~8& 1& 21 & 1&~~7 & 5+2?& 13& 1?&~~7\\
 Quintuplet$^{**}$ &14&14&84& 9&~~5& 1& 14 & 0&~~5 & 5+2?&~~9& 1?&~~5\\
  Arches$^a$        &17&13&15& 0& 17& -&-         & 0& 15 &    -&- &2+5?&17\\
  Central Parsec$^b$&42&30&40&24& 18& -&-         & 3& 12 &  10?& 24& 2& 18\\
  Westerlund\,1$^c$ &24&24&57& 8& 16&10& 24 &12&~~4 &   6?&~~8&3+6?&16\\
 \hline
  Milky Way$^d$     &63& -&- &38& 25& -&-         &33& 26 &  19 & 38& 6&25\\ 
\hline \hline
  \end{tabular}
 \tablefoot{Numbers of O stars are lower
    limits from \citet[Arches]{Martins+2008}, \citet[Central
    Parsec]{Paumard+2006}, \citet[Wd\,1]{Negueruela+2010}; WR numbers
    are based on \citet{vdH2006} unless otherwise indicated. WC stars
    showing dust in their spectra (WCd or WCLd classification) are
    considered binary candidates.
    Note that the number stars listed in the different columns for each
clusters might slightly vary in dependence of the reference and the
therein used subsample of stars (see text for details).
 \begin{list}{}{}
  \item[$^*$]{WR stars from \citet{vdH2006} corrected for the WR stars
  newly identified in \citet{Liermann+2009}: LHO\,110 - WN9h, LHO\,76 and
 LHO\,79 both WC9d; number of RSGs estimated from the corrected sample
  of late-type stars; five WCd stars resolved as pinwheel binaries by
  \citet{Tuthill+2006}, two further candidates based on WCd classification.}
  \item[$^{**}$]{Numbers limited to the LHO field of view of about
      40\,\arcsec $\times$ 40\,\arcsec covering the center of the
      Quintuplet cluster.}
  \item[$^a$]{Classification as WNL or WNE from \citet{Cotera+1999} and
      \citet{Martins+2008}; two WN binaries from non-thermal, variable
      radio plus X-ray detection, further 5 candidates based on either
      variable radio flux or X-ray detection \citep{vdH2006}.}
  \item[$^b$]{Classification as WNL or WNE  from
      \citet{Paumard+2006,Martins+2008}; two Of/WN9 stars detected
      with radial velocity variations \citep{Martins+2006}.}
  \item[$^c$]{$N_\mathrm{RSG}$ includes four RSGs by
      \citet{Mengel+Tacconi2007} and six yellow hypergiants by
      \citet{Clark+2005}; 
      classification as WNL or WNE and WR binaries taken from
      \citet{Crowther+2006}.}  
  \item[$^d$]{Numbers from \citet{vdH2001},
      Table\,29. $N_\mathrm{WNE}$:$N_\mathrm{WNL}$ ratio from \citet{HGL2006}.}
 \end{list}
}
  \end{center} 
\end{table*}
%________________________________________________________________

Concerning their position in the HRD below the 9\,$M_\odot$ track and
off the main sequence, the late-type KM stars would have to be considered
evolved low- and intermediate stars. This is confirmed by the presence
of $^{13}$CO absorption in their spectra which led to the classification as
(super-) giants (LHO catalog). However, a few late-type stars
extend to the regime of massive stars touching the 9 to 15\,$M_\odot$ tracks.
Their potential status as red supergiant (RSG) will be discussed separately
(see below). 

In Fig.\,\ref{fig:cluster-hrd+isochrones} the HRD with
isochrones compiled by \citet[based on Geneva stellar
evolution models]{Lejeune+2001} is shown for different cluster ages. 
For comparison we show the isochrones from \citet[``Padova isochrones'' in the
following]{Girardi+2002} in Fig.\,\ref{fig:cluster-hrd+isochrones2}.
In both cases, the assumed cluster main sequence of 
OB stars corresponds well with the 4\,Myr isochrone with the 3 and
5\,Myr isochrone forming an ``age envelope''. 

In case of the WN stars, two stars (WR\,102d and WR\,102i) seem to
follow the Geneva 2\,Myr isochrone,
while the 2\,Myr isochrone from Padova is below any WN star position
(Fig.\,\ref{fig:cluster-hrd+isochrones2}). In addition, both of these
isochrones don't extend far enough into the cooler region of the HRD
to cover all WN stars.
Interestingly, the Padova isochrones for 3, 4 and 5\,Myr loop backwards
in the hotter region of the HRD, and especially the 3\,Myr isochrone
corresponds very well to the position of {\it all} the evolved WN stars.
A detailed comparison of the WN stellar parameters with the
Geneva stellar evolution models \citep{Meynet+Maeder2003} revealed
ages of about 2.1 to 3.6\,Myr for the Quintuplet WN stars \citep{Liermann+2010,
  Liermann+2011Liege}; an age of 3\,Myr seems to be confirmed
by the Padova isochrone. The presence of WC stars in the Quintuplet
cluster speaks neither in favor of a much older nor a younger cluster
age, since they may originate from binary evolution as explained above.

The same argument holds for the luminous blue variables (LBVs)
in the Quintuplet, the Pistol star and qF\,362. From high-mass single star
evolution the LBV phase can be estimated to start at about
4\,Myr, which was also used as age criterion for the Quintuplet
\citep{FMM99, Geballe+2000}. However, recent interferometric studies
indicate that the Pistol star, just like the LBV prototype
$\eta$\,Car, might be a binary and thus evolved through the binary
channel \citep{Martayan+2011}.

A similar spread in age between O stars and WR stars was reported by
\citet{Martins+2008} for the Arches cluster.  They found ages of 2 to
3\,Myr for the most luminous WN stars, while the O stars cover a range
of 2 to 4\,Myr. They conclude that this might be due to the fact that
the most massive stars have formed last in the cluster, to prevent
their feedback on the ongoing star formation. In the end they
determine a cluster age of $2.5 \pm 0.5\,$Myr.  Similar arguments can
be applied for the the difference in age of the OB and WN stars in the
Quintuplet. Thus we conclude that a cluster age of $3.5 \pm 0.5\,$Myr
is likely.

Both sets of isochrones suggest that the KM stars, if they were
cluster members, would need about 15\,Myr to evolve to RSGs for
$M_\mathrm{ini} > 9\,M_\odot$ and more than 30\,Myr for $M_\mathrm{ini} <
9\,M_\odot$ to become red giants, respectively. The detected $^{13}$CO
gives spectroscopic evidence that these stars are evolved giants or
supergiants \citep{Liermann+2009}. It appears that for the majority of
these stars it might not be justified to assume that they are located at
a distance of 8\,kpc. They can be foreground objects and/or physically
unrelated to the Quintuplet cluster. This conclusion seems to be
supported by the radial velocities of most of these stars which are off
the limits of the corrected sample for the cluster. However, we are left
with a group of eight evolved KM stars that might be physical 
members of the Quintuplet (see below).

A priori the coexistence of RSGs, main sequence OB stars, and WR stars
is not expected under the assumption of a coeval evolution. However,
\citet{Hunter+2000} discuss the possibility of an extended star
formation event lasting over a few million years, which could explain
the simultaneous presence of WR stars and RSGs. We find an age range
of about 4\,Myr (OB stars) to 15 to 30\,Myr (KM stars) for the two
populations. Could such a spread in age be called an ``extended'' star
formation event? Or is it more likely that one star formation event was
responsible for the WR and OB stars while the evolved KM stars
might represent an older population that is unrelated to the cluster?

\subsection{Number ratios and the binary effect}
In Table\,\ref{tab:number-ratios} we list stellar number ratios for
some prominent clusters and the Milky Way in total.
These numbers are only approximates. The
ratios can suffer from systematic observational errors, incomplete
samples used in the different references, or from unknown/unconfirmed
binary stars in the samples.
For example, among the general results of \citet{vdH2001} for the Galactic
WR star population, they give a $N_\mathrm{WC}/N_\mathrm{WN}$ ratio of
1.5 and a binary frequency for the WR stars of 39\%. These numbers are
based on a stellar subsample limited to WR stars within 2.5\,kpc;
presumably a complete sample. On the other hand, their total catalog
numbers would result in a different $N_\mathrm{WC}/N_\mathrm{WN}$
ratio of 0.7, which might lead to biased estimates of the
WR subclasses.
\citet{HGL2006} selected only putatevely single stars from the WR star
catalog \citep{vdH2001} for their study of the ratio between
different spectral subclasses. The binarity status of invidual objects
may have been revised in the meantime, and thus affect the number ratios. 

For the Arches and the Central Parsec cluster the situation is similar, depending
on the study, their field of view or the available stellar subsample, the
numbers in the different columns of Table\,\ref{tab:number-ratios} may differ
from those listed in \citet{vdH2006}.

With these uncertainties in mind, we compare the number ratios 
$N_\mathrm{WR}/N_\mathrm{O}$ and $N_\mathrm{WC}/N_\mathrm{WN}$ with
the predictions of population synthesis models as presented
  in the figures by \citet[{\sc starburst99}]{Leitherer+1999,
  Vazquez-Leitherer2005}. The models account for solar metallicity and
single star evolution. From \citet[(Fig.\,39 to
  42)]{Leitherer+1999}, we find lower limits for the age of the
Quintuplet cluster of 3.3 to 3.6\,Myr in the scenario of instantaneous
burst star formation. This agrees very well with the above HRD and
isochrones.  Although we cannot exclude a past supernova in the
Quintuplet, the determined age would also be conform with the
predictions of \citet[(Fig.43)]{Leitherer+1999} in terms of
first supernovae taking place only at about 3.5\,Myr in the star burst
scenarios.

One issue with the Quintuplet stellar population is obvious: 
All WN-type stars in the Quintuplet have been classified as WNL,
i.\,e. they still contain hydrogen in their atmosphere
\citep{Liermann+2009,Liermann+2010}. WNE stars (hydrogen-free WN
stars) are significantly missing, although the number ratio
$N_\mathrm{WC}/N_\mathrm{WR}$ is compatible with the value found by
\citet{vdH2001} for the Milky Way. From 
the regular Milky Way WR population we would expect a slightly 
higher number of WNE stars than WNL stars \citep{HGL2006}. 
WR\,102c, which lies outside the LHO field of view in the Sickle nebula,
was classified by \citet{FMM99} as WN6 type star and thus belongs to
an ``early'' subtype (WNE). But \citet{Barniske+2008} found from
comparison with model spectra that the stellar atmosphere still
contains up to 20\,\% hydrogen, making it a ``WNL'' star in the
evolutionary sense.
Interestingly, WNE stars seem to be missing in the other clusters in
the Galactic center region as well \citep[see][ and
Table\,\ref{tab:number-ratios}]{Liermann+2010}. 
From the Geneva models we know that hydrogen burning lifetimes and WNL
lifetimes are increased when the effects of rotation are included,
while WNE lifetimes are hardly affected. This might skew the number
ratios in favor of WNL stars.
\cite{Liermann+2010} noticed that evolutionary models without rotation
represent the Quintuplet WNL stars better than those with rotation. 
This might imply that the initial rotational velocity of
300\,km/s$^{-1}$ in the models is too much. Or, as pointed out by
\citet{Vanbeveren+2007}, the mismatch between evolution tracks and
observations for massive stars may be due to inappropriate assumptions
of the mass loss in the RSG stage. 

So far we have only considered single star evolution, but what about
binaries? The most promising indications for binarity seem to be
present among the WR stars, so we will focus on those in the
following. In the Milky Way about a
quarter of all WN stars and half of all WC stars are in binaries
\citep[see also Table\,\ref{tab:number-ratios}]{vdH2001}. As 
noted by several authors \citep[e.\,g.][]{vdH2006, Crowther+2006,
  Liermann+2009}, $K$-band spectra that show an increasing continuum
with wavelength (IR excess) can be interpreted as being colliding wind
binaries in which dust is formed \citep{Williams+1987}. The majority
of WC stars in current catalogs and surveys are binary 
candidates on this ``dust'' argument alone.

With no time series of spectral observations available, the LHO
spectra allow only indirect conclusions about binarity to be
drawn. Among the WC stars in the Quintuplet are nine stars classified
as WCd, i.\,e. dust producing colliding wind binary candidates. Five of
those, the enigmatic Quintuplet pinwheels, can be considered confirmed
binaries as they were partially or fully spatially resolved by
interferometry by \citet{Tuthill+2006}. Two further WC stars, LHO\,76
and 79, have been classified in the LHO catalog as potential binaries
on the basis of a ``dusty'' spectrum. So, the number of WC binaries
might even be higher.

Among five WN stars in the LHO catalog, we find only one binary
candidate: LHO\,110. \citet{Liermann+2010} found
small absorption lines in the spectrum which might belong to an O-type
companion. This is corroborated by its high measured radio flux.
One further WC and one WN star are listed by \citet{vdH2006} as binary
candidate. But for these, WR\,102f (WC8+?) and WR\,102i (WN9+?), the LHO
spectra look unsuspicious, which is why the classification as binary
candidates was dropped.
The number of unconfirmed binaries makes it difficult to compare the
binary fraction with other Galactic center clusters and the Milky
Way (see Table\,\ref{tab:number-ratios}). But it seems that the binary
fraction of the Quintuplet WN and WC stars resembles roughly the
one for the Milky Way in general.
 
\citet{Vanbeveren+1998} argue that taking binaries into
account is required to get correct $N_\mathrm{WR}/N_\mathrm{O}$ and
$N_\mathrm{WC}/N_\mathrm{WN}$ ratios. Their population synthesis
suggests that, especially for the inner Milky Way, a 
$N_\mathrm{WR}/N_\mathrm{O}$ ratio of the order of 0.2, a value
similar to the one found for the Quintuplet, cannot be explained with
a continuous star formation scenario. It could rather be explained
with an enhanced star formation or star burst. In addition, according
to those models a star burst that includes binaries encounters the
effect of rejuvenation, i.\,e. mass transfer which takes place in 
massive close binaries leads to a population of young O-type stars
mimicking a much younger age. This is supposed to happen after
about 4\,Myr, and would manifest in a group of blue young stars
above the main sequence turn-off of the main population. 
We don't see a prominent group of such O-type stars in the HRD of the
Quintuplet, which eliminates a cluster age of $\ge
6$\,Myr according to \citet[Fig.\,11~B]{Vanbeveren+1998}. Given the
error bars in our empirical $T_\mathrm{eff}$ and $\log{L}$
(Fig.\,\ref{fig:cluster-hrd+tracks}), and comparing this to the spread
in \citet[Fig.\,11~A]{Vanbeveren+1998}, we can furthermore limit the
cluster age to $\le 4$\,Myr. This is within the range we estimate from
the main sequence of OB and WR stars and the isochrones.

From the good agreement with the overall numbers ratios
\citep{vdH2001} and the age limits inferred from single and/or binary star
evolution \citep{Vanbeveren+1998, Leitherer+1999} it seems secure to
conclude that the formation of the Quintuplet cluster most likely
needed a star burst event about $3.5 \pm 0.5$\,Myr ago.

\subsection{Late-type stars in the Quintuplet?}

%________________________________________________________________
%
\begin{figure}
         \center{
\includegraphics[width=0.95\columnwidth]{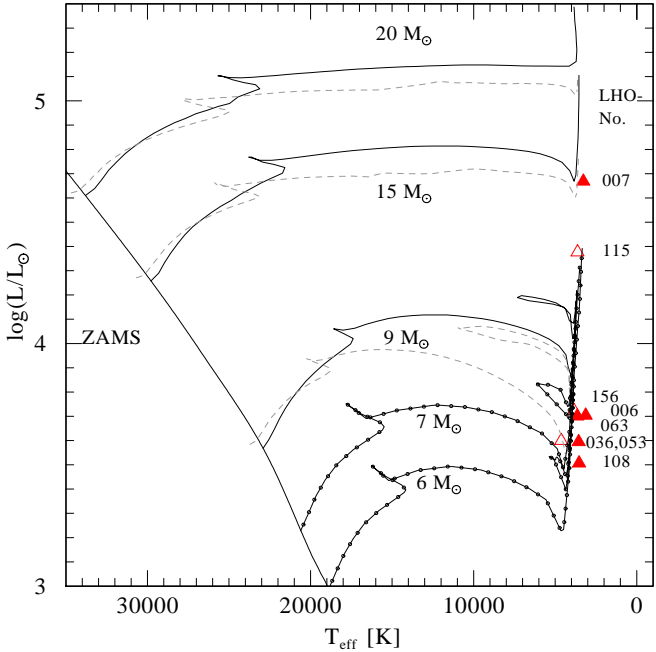}}
\caption{Detail of Fig.\,\ref{fig:cluster-hrd+tracks} with focus on
  the late-type stars with $\log{(L/L_\odot) > 3.4}$ that
  might be Quintuplet RSG candidates. 
  Tracks show stellar evolution models from \citet{Meynet+Maeder2003}
  with and without rotation (solid and dashed lines, respectively) and
  from \citet[solid line with circles for $M_\mathrm{ini}= 6$ and $7\,M_\odot$]{Bressan+1993}.}
      \label{fig:hrd-RSGs}
\end{figure}
%________________________________________________________________
%

According to \citet{Vanbeveren+1998} almost one third of all Galactic
open cluster and stellar aggregates host WR stars and RSGs
simultaneously. In the majority of them the total numbers are
rather small, of the order of one or two, and in all cases the number
of WR stars exceeds those of RSGs. 

In terms of abundance of these special stars, Westerlund\,1 (Wd\,1) is
the most prominent Galactic open cluster known to simultaneously
contain evolved early-type stars, such as luminous blue variables
(LBVs), yellow hypergiants (YHGs), WR stars, and RSGs. 
Based on single star evolution, \citet{Crowther+2006} concluded from
an number ratio of $N_\mathrm{RSG+YHG}/N_\mathrm{WR}=$ 8:24 a cluster
age of 4.5-5\,Myr for Wd\,1.  
\citet{Mengel+Tacconi2007} list four RSGs plus six YHGs found by
\citet{Clark+2005} which changes 
the number ratio slightly (see Table\,\ref{tab:number-ratios}). 
\citet{Vanbeveren+1998} argue on basis of single star lifetimes
that WRs and RSGs can be present simultaneously only in a very
limited time span of 4 to 5\,Myr. Too early no RSG would be present and
too late the WR stars would be already gone.

We already noted that in the Geneva models with rotation the WNL lifetime
is increased. In addition, \citet{Hirschi+2004} found that in the rotating
models the RSG phase could be reached even before helium ignites in
the stars, much earlier than in non-rotating models, and that the RSGs
become more luminous. Similar effects were found for the WNL stars.
Both WNL and RSGs can appear spectra-wise as evolved stars,
while they are still core-hydrogen burning. This might twist
the number ratios towards more WNL than WNE stars plus the
simultaneous presence of RSGs even at young cluster ages. 

For the Quintuplet cluster, eight late-type stars 
in the regime of $\log{(L/L_\odot)} > 3.4$ 
are found (see Fig.\,\ref{fig:hrd-RSGs}).
Including stellar evolution tracks from \citet{Bressan+1993}, 
most stars below $\log{(L/L_\odot)} < 4$ could also
be regarded as intermediate-mass stars (tracks with $M_\mathrm{ini}
\approx 6$ to 7\,$M_\odot$). 
However, only five out of eight of these stars are from the
corrected sample of late-type stars, i.\,e. potential cluster members,
due to their $RV$. This would result in
a number ratio $N_\mathrm{RSG+YHG}/N_\mathrm{WR}$ of 5:21 similar  
as for Wd\,1 (see Table\,\ref{tab:number-ratios}) and could imply
that the Quintuplet cluster is slightly older than just 4\,Myr.

The stars LHO\,006, LHO\,036, LHO\,053, and
LHO\,063 have not yet been studied in detail. 
The remaining stars are discussed individually in the following: 

\paragraph{LHO\,007} - alias GMM\,7 (Q\,7), qF\,192\\
\citet{GMM90} mentioned this prominent and bright source as a
late-type star in the Quintuplet region.
\citet{MGM1994} found the star to be an M-type supergiant with $K
= 7.36\,$mag, and \citet{GMC99} consider the star to be not variable.
The stellar parameters from a detailed study by \citet{Blum+2003} are
compared with our results in Table\,\ref{tab:LHO007}. Similar $K$-band
magnitudes but slightly different effective temperatures put the star at
the highest luminosity within our group of late-type stars. 
A number of studies also focused on the chemical composition of the
star \citep[e.\,g.][]{Ramirez+2003, Cunha+2007, Davies+2009}, finding
iron abundances to be nearly consistent with solar values but slightly
enhanced $\alpha$ elements. Most authors agree that
LHO\,007 is an evolved star with an initial mass likely between 15 and
20\,$M_\odot$ depending on the stellar evolution model. From our
results we find that all stellar evolution models (with and without
rotation, from Geneva and Padova) imply an initial mass of
15\,$M_\odot$ (Fig.\,\ref{fig:hrd-RSGs}). 
The star's membership to the Quintuplet cluster was debated in terms
of extinction issues \citep{GMM90,MGM1994, GMC99}.
According to its $RV = 124\,$km\,s$^{-1}$ it may belong to the
Quintuplet cluster.

%________________________________________________________________
\begin{table}[t]
  \caption{Comparison of stellar parameters for LHO\,007.}
  \label{tab:LHO007}
  \begin{center}
  \begin{tabular}{lcc} \hline \hline
  \noalign{\smallskip}
                        & \citet{Blum+2003} & this work\\
  \noalign{\smallskip} \hline
  \noalign{\smallskip}
   Spectral type        & M\,\sc{i}         & M6\,\sc{i} \\
   $T_\mathrm{eff}$ [K]  & 3570              & 3274       \\
   $K_\mathrm{s}$ [mag]  & 7.3               & 7.6        \\
   BC [mag]             & 2.6               & 3.09       \\
   $M_K$ [mag]          &$-$10.22             & $-$10.02     \\
   $M_\mathrm{bol}$ [mag]&$-$7.6               & $-$6.93      \\ 
   $\log{(L/L_\odot)}$   & 4.94              & 4.67       \\
\hline \hline
  \end{tabular}
  \end{center} 
\end{table}
%________________________________________________________________

\paragraph{LHO\,108} - alias [NWS90] A, [GMC99] D\,322, qF\,269\\
This star is considered to be a foreground star by \citet{NWS90}.
In contrast, \citet{GMC99} list the star in the group of probable
cluster members. Its $RV = 127\,$km\,s$^{-1}$ puts it in our
corrected sample. \citet{FMM99} list the star in their Quintuplet
sample, with a spectral classification as OBI and a luminosity of
$\log{(L/L_\odot)} = 5.54$. But the LHO spectrum definitely shows CO bands
classifying it as late-type star (see LHO catalog spectra). Thus we
derive a much smaller luminosity of $\log{(L/L_\odot)} = 3.51$. It
places the star at the lower end of the group in vicinity
to the non-rotating 9\,$M_\odot$ track from Geneva or the 6 to
7\,$M_\odot$ tracks from Padova (see Fig.\,\ref{fig:hrd-RSGs}). 

\paragraph{LHO\,115} - alias GMM\,5 (Q\,5), MGM\,5-5, qF\,270N\\
This star is one of the five from which the Quintuplet cluster derived
its name. Its $K$-band magnitude is 8.71\,mag \citep{GMM90}. 
Later works give a range of values:
$K = 9.1\,$mag \citep{MGM92}, 
$K = 8.78\,$mag \citep{MGM1994}, 
$K = 7.81\,$mag \citep[spectral type ``late'']{FMM99}. 
\citet{GMC99} list the star in their sample of long-period variables
in the Quintuplet with an average $K = 8.62\,$mag and a 680\,d
photometric variability of 1.5\,mag. 
They attribute the variability to the star being a Mira variable and
argue on basis of stellar evolution that the star would then be too old to
be a Quintuplet member. However, according to its photometric colors
\citep{MGM1994} and its apparent magnitude being in the range expected
for Miras in the GC region \citep{Glass+1999}, the cluster membership
might still be likely.

From our $K= 8.6\,$mag we derive a luminosity of
$\log{(L/L_\odot) = 4.38}$, which makes the star the second most
luminous in this group of RSG candidates.
Its position in the HRD (Fig.\,\ref{fig:hrd-RSGs}) puts it at the
stellar evolution track for a star with an initial mass of
9\,$M_\odot$. With the $RV = 137\,$km\,s$^{-1}$
it dropped out of the corrected sample, but still is within the 3\,$\sigma$
interval of possible cluster membership. 

\paragraph{LHO\,156} - alias [GMC99] D\,307\\
\citet{GMC99} list this star with $K= 10.83\,$mag in the sample of
probable Quintuplet members and show a slightly
variable light curve. However, no firm conclusion about a possible 
variability is drawn. We find $K= 10.4\,$mag in the LHO catalog. It's
derived luminosity 
$\log{(L/L_\odot) = 3.72}$ puts it on the non-rotating 9\,$M_\odot$
track from Geneva or the tracks from Padua with 6 to 7\,$M_\odot$
initial mass. However, with $RV = 169\,$km\,s$^{-1}$ the star is
outside the limits for the corrected sample and even outside of the
3\,$\sigma$ cluster $RV$. 

\vspace{.2cm}\noindent
Summarizing, there are a few KM stars in the Quintuplet field
which share the radial velocity of the cluster. One of them, LHO\,007
alias Q\,7, has supergiant luminosity. Other ones are red-giant branch
stars, assuming Galactic center distance. One can exclude that they are
foreground KM dwarfs, because of their radial velocity and because of
the $^{13}$CO signature in their spectra that is indicative for giants
and supergiants. It would be very puzzling if these relatively old
($>$30\,Myr) stars would be gravitationally bound to the Quintuplet cluster.
Otherwise, they may belong to an old field population in the Galactic
center region.

It seems that the assessment of the cluster membership, e.\,g. from
proper motion studies 
\citep{Hussmann+2011}, is needed 
to settle the question if older stars belong to the Quintuplet. 
As discussed in the previous section, binary evolution
\citep{Vanbeveren+1998} and stellar evolution with rotation
\citep{Hirschi+2004} may explain the coexistence of RSG and WR
stars in a cluster of about 4\,Myr, as the likely age of the Quintuplet.
Alternatively, if none of these scenarios works, the conclusion would
become inevitable that the cluster contains a second, older population
of stars.

%________________________________________________________________
%
   \begin{figure}
   \centering
   \includegraphics[width=\columnwidth]{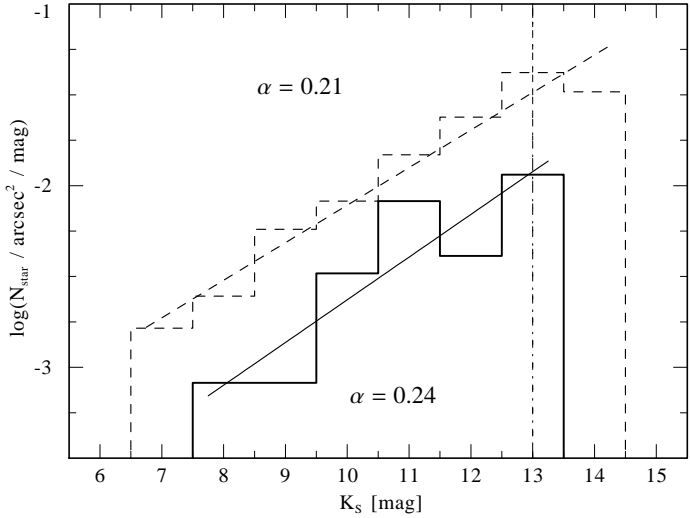}
   \caption{Luminosity function of the Quintuplet cluster
     for all stars in the LHO catalog (dashed line) and
     for the corrected sample (solid line). The photometric
     completeness at $K_\textrm{s}=13\,$mag is indicated by the
     vertical dashed line. The linear fits to both distributions are
     indicated by the slope coefficient $\alpha$.}
               \label{fig:LF}
   \end{figure}
%________________________________________________________________
%

\subsection{Luminosity function and mass function}
To further exploit the LHO data of the Quintuplet cluster, we
establish the luminosity function by counting the number of stars per
magnitude bin in our total cluster field of view; in addition we
apply a normalization factor to account for the cluster area.
The resulting distribution is shown in Fig.\,\ref{fig:LF}.
A power law with an exponent $\alpha = 0.24 \pm 0.06$ can be
fitted to the distribution of the corrected sample.
\citet{Figer+1999}
argue that such kind of shallow distribution is expected for young
coeval clusters. \citet{Harayama+2008} found for example $\alpha =
0.27$ for the young massive star-forming region NGC\,3606.

%________________________________________________________________
%
   \begin{figure}%[!hb]
   \centering
   \includegraphics[width=\columnwidth]{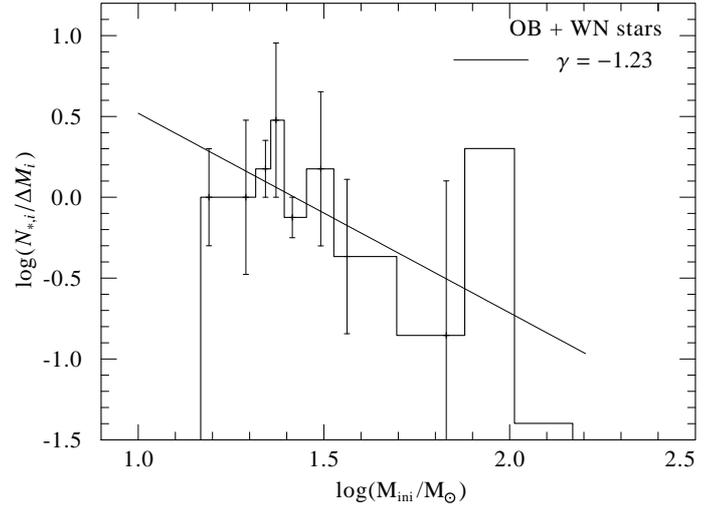}
   \caption{IMF for the early-type OB stars in the
     corrected sample of the LHO catalog plus the WN stars.
     The distribution shown has variable mass range bins with a
     constant number of stars per bin to avoid numerical biases according to 
     \citet{MaizApellaniz-Ubeda2005}; the linear fit gives a slope coefficient as
     indicated in the plot. 
}
               \label{fig:IMF}
   \end{figure}
%________________________________________________________________
%

For a young coeval cluster it is expected that
the mass distribution in the stellar population still resembles the
initial mass distribution. 
However, the evolved WR stars (WN and WC subtypes) and the presence of
RSG candidates clearly show that some stars have already undergone
significant mass-loss.
On the other hand, we found that the majority of early-type stars
form a main sequence around a 4\,Myr isochrone. 
Thus, we assume the OB stars to have formed during one
star formation event and omit the evolved WR stars and
late-type stars for this moment from the further discussion.

From the Geneva 4\,Myr isochrone a mass-luminosity relation
is constructed and 
applied. Our completeness limit of $K_\mathrm{s} = 13\,$mag translates
to a minimum mass of $M_\mathrm{ini} = 10\,M_\odot$ for an OB star.
Thus we obtain initial masses in the range
of 10 to 78\,$M_\odot$, with most stars concentrating in the 10
to 30\,$M_\odot$ range and much fewer stars for higher mass bins. 
The error bars are estimated on the basis of the uncertainty of the
luminosity propagating through to the derived masses.

To determine the IMF we follow \citet{MaizApellaniz-Ubeda2005} and use bins
with a variable mass range but a constant number of stars per bin to
be robust against numerical biases. In Fig.\,\ref{fig:IMF} we show the
resulting distribution for the OB stars in the corrected sample plus
the WN stars with their masses as determined by \citet{Liermann+2011Liege}.
A linear fit of the form $\log{(\mathrm{d}N/\mathrm{d}M)} \sim \gamma \times \log{M}$ gives a power law with $\gamma = -1.23 \pm 0.51$. The
fit for the OB stars only gives a slightly steeper result of $\gamma
= -1.66 \pm 0.51$ as the WN stars basically populate the very high-mass bins.
Both scenarios suggest a slightly top-heavy IMF in comparison to the
canonical Salpeter-IMF \citep[$\gamma$ = -2.35]{Salpeter1955}. 

Tests with the Padova 4\,Myr isochrone give similar results.

Top-heavy IMFs are discussed for massive stellar cluster
throughout the Galaxy and especially for the Galactic
center region. For example, \citet{Maness+2007} find $\gamma
= -0.85$ for the Central Parsec cluster assuming continuous star
formation, albeit \citet{Paumard+2006} favor a star burst scenario for
this cluster. The Arches cluster was analyzed by \citet{Stolte+2005}
who discuss a flat present-day mass function with $\gamma = -1.9$ to
$-2.1$, but \citet{Portegies-Zwart+2002} do not exclude a regular
Salpeter-IMF for that cluster when mass segregation is taken into account.
Our results on the Quintuplet cluster seem to be in
  agreement with the findings of a recent study by
  \citet{Hussmann+2011}.

\section{Summary}

In this paper we have used the $K$-band spectra of the LHO catalog to
obtain stellar temperatures and luminosities for all early- and
late-type stars in the Quintuplet cluster. Furthermore, we used the
LHO catalog to obtain number ratios for the different stellar
subclasses. As result we present a
cluster HRD that is compared with stellar evolution tracks and
isochrones; the main findings are the following:

\begin{enumerate} 

\item The early-type OB stars in the Quintuplet are high-mass stars
  with initial masses $>$ 8\,$M_\odot$. They form a main sequence
  population of massive stars around an isochrone of about 4\,million
  years.

\item For the WN stars we determine an age of 3\,million years from
  the isochrones which is consistent with the ages found by
  \citet{Liermann+2010}.  A possible interpretation for the
  slight difference to the age of the OB stars is that the most massive
  stars might have formed late in the cluster formation process.

\item
Alternatively, if all WR stars in the Quintuplet evolved as secondary
stars in binary systems, they might have been rejuvenated by mass
transfer while being originally older. 

\item Considering different number ratios
  (e.\,g. $N_\mathrm{WR}/N_\mathrm{O}$, $N_\mathrm{WC}/N_\mathrm{WN}$)
  in the Quintuplet we find best agreement from comparison with
  population synthesis models for a star burst that formed the
  Quintuplet cluster about 3.5$\pm$0.5\,million years ago.

\item The late-type KM stars need much longer to evolve to their present
stage. According to their radial velocities the majority of them cannot
be considered as cluster members. Five of the KM stars, however, share
the radial velocity of the Quintuplet cluster. One of them (LHO\,007
alias Q\,7) is a supergiant, the others are less luminous. They may
represent an older field population in the Galactic center region.

\item If LHO\,007 really is a physical member of the Quintuplet cluster,
this would have interesting implications. With standard single-star
evolution, the simultaneous presence of WR stars and a RSG of initially
about 15\,$M_\odot$ is only expected for clusters of much older age. 

\item
Recent stellar evolution models which take the effects of
rotation into account, as well as considering
binary evolution channels, may help to explain the coexistence
of WR stars and RSGs in a cluster of only about 4\,million years.

\item We derive the luminosity function of the Quintuplet cluster with
  a power-law slope of $\alpha = 0.24 \pm 0.06$. This is consistent
  with a young coeval cluster and confirms the star burst scenario for the
  formation of the Quintuplet.

\item For the subgroup of the high-mass stars we derive a
  mass-luminosity relation from the 4\,million year isochrone and
  apply it to obtain the IMF. 
  The result is a power law with a slope that suggest a slightly
  top-heavy IMF compared to the Salpeter-IMF. If we include the WN
  stars the effect is more pronounced.
\end{enumerate}

In summary, the Quintuplet cluster seems to have an intermediate
age of 3.5$\pm$0.5\,million years compared to the other Galactic
center clusters, the Arches 
\citep[2.5$\pm$0.5\,million years,][]{Martins+2008} and the Central cluster
\citep[6$\pm$2\,million years,][]{Paumard+2006}. As recently discussed by
\citet{Boeker+2008}, an age gradient of young stellar clusters in
extragalactic nuclei can be explained in a scenario of 
almost continuous star formation by over-density regions in a
circum-nuclear ring (CNR) in a galaxy. This would lead to a burst-like release
of stellar clusters in regular time steps like ``pearls on a string'' in the
central zone. From radio data we
know the existence of the Milky Way's CNR and the active 
high-mass star formation region Sgr~B2
\citep[e.\,g.][]{Morris-Serabyn1996}. Thus, if the star formation in
the inner Milky Way is interpreted in the way described
above, this would explain the observed age sequence in the three
Galactic center clusters.

\begin{acknowledgements}
  We thank the anonymous referee for the suggestions to improve the
  manuscript.
  The authors are very thankful to B.~Davies who kindly shared his
  knowledge about red supergiant stars.
  Furthermore discussions with A.~Stolte and B.~Hu\ss\-mann
  about the cluster membership and cluster age were of great value.
  We thank J.~Groh for comments that helped to improve the manuscript.
  The work was supported in its early phase by the Deutsche
  Forschungsgemeinschaft (DFG) with grant HA 1455/19. 
\end{acknowledgements}

\bibliographystyle{aa}

\end{document}